# About density functional theory interpretation

Kirill Koshelev

E-mail: kirvkosh@gmail.com

**Abstract**

Two forms of relativistic density functional are derived from Dirac equation. Based on their structure analysis model of split electron is proposed. In this model electric charge and mass of electron behave like two point-like particles. It is shown that two electrons obeying this model cannot occupy the same quantum state. Empirical verification of the model is discussed.

**Introduction**

A few decades ago that time a brand new approach to explain the properties of quantum systems was proposed, namely density functional theory. Starting with famous Hohenberg-Kohn theorem proved for nonrelativistic case and later generalized for relativistic systems Ref. (1) the method became one of the main tools for calculation of condensed matter electronic structure. The idea behind the density functional theory is an existence of relationship between energy of a system of electrons and their probability density distribution expressed in terms of density functional. Surprisingly, up to now the exact form of a corresponding functional is not found. Today we can only operate some approximate and mostly semi-empirical forms of the functional in order to make estimations. Despite of the absence of the rigorous definition of said functional the whole approach is rather regarded as a mathematical trick without any deep physical interpretation behind it.

In present paper we analyze formulas for relativistic density functional received earlier Ref. (2) and give interpretation to explain the very fact of their existence. In order to do this model of split electron is proposed, where an electron obeying Dirac equation possesses two particle structure where the first particle carries all the mass of electron and the second one carries all the charge of electron. This model explains why formal velocity of electron received from Dirac equation amounts to speed of light in vacuum. The other immediate consequence of the model is a deviation of electric charge distribution in hydrogen-like ion in nonrelativistic limit from the prediction of a Schrödinger equation. This outcome can be used for experimental verification of the model. Throughout in the paper we use atomic system of units.

**Dirac equation and density functional**

Let's consider an electron in hydrogen-like ion obeying relativistic Dirac equation. Hamiltonian $H$ for relativistic electron moving in the Coulomb potential can be chosen in the following form Ref. (3)

$$H = c(\vec{\alpha} \cdot \vec{p}) + V + mc^2\beta \qquad (1)$$

where $V = -\frac{e^2 Z}{r}$ is Coulomb potential of a point-like nucleus, $\vec{p}$ is a momentum operator of electron, $m$ and $c$ are electron mass and speed of light in vacuum constants respectively, and finally $\vec{\alpha}$ and $\beta$ are set of Dirac $4 \times 4$ matrixes

$$\vec{\alpha} = \begin{pmatrix} 0 & \vec{\sigma} \\ \vec{\sigma} & 0 \end{pmatrix} \quad \beta = \begin{pmatrix} I & 0 \\ 0 & -I \end{pmatrix} \tag{2}$$

To find out eigen functions and corresponding energies one solves the equation

$$H\Psi = E\Psi \tag{3}$$

where $\Psi = (\Psi(1), \Psi(2), \Psi(3), \Psi(4))^T$ is a four component wave function and $E$ is associated eigen energy. It is demonstrated in Ref. (4) that application of the virial theorem to Eq. (3) produces the following formula

$$E = mc^2 \langle \Psi | \beta | \Psi \rangle \tag{4}$$

and analogously the virial theorem applied to the Eq. (3) with squared Hamiltonian Ref. (2) yields

$$E^2 = m^2 c^4 + mc^2 \langle \Psi | V\beta | \Psi \rangle \tag{5}$$

Also in Ref. (2) was shown that the following formulas are valid for the wave function

$$E_b = 2 \int \rho_e \, d\tau \tag{6}$$

and

$$mc^2 + \frac{1}{2} E_b = \int \rho_m \, d\tau \tag{7}$$

where $E_b = E - mc^2$ is a binding energy for the electron, and $\rho_m = mc^2(|\Psi(1)|^2 + |\Psi(2)|^2)$ and $\rho_e = -mc^2(|\Psi(3)|^2 + |\Psi(4)|^2)$. Expressing wave function in Eq. (5) via densities $\rho_m$ and $\rho_e$ and applying Eq. (6) and (7) we come to the formula

$$E^2 = m^2 c^4 + \frac{E+mc^2}{2} x + \frac{E-mc^2}{2} y \tag{8}$$

where $x = \int V \tilde{\rho}_m \, d\tau$ and $y = \int V \tilde{\rho}_e \, d\tau$, and $\tilde{\rho}_m$ and $\tilde{\rho}_e$ are normalized densities $\rho_m$ and $\rho_e$ so

$$\int \tilde{\rho}_m \, d\tau = \int \tilde{\rho}_e \, d\tau = 1 \tag{9}$$

As it is shown in Ref. (2) formulae Eq. (4) (that is a density functional) can be easily generalized for multi-electron case. Apparently the virial theorem plays critical role to infer the expressions for density functional. Nevertheless keeping in mind that density functional exists for great variety of potentials beyond the application of virial theorem we can make a conclusion that there should be a fundamental reason for existence of such functional in general case.

**Split model of electron and Pauli Exclusion Principle**

Formula Eq. (8) provides with opportunity to find the energy of relativistic electron subject to two distributions $\tilde{\rho}_m$ and $\tilde{\rho}_e$ are found. The distribution $\tilde{\rho}_m$ in nonrelativistic limit turns into electron density constructed from the wave function obeying one electron Schrödinger equation. Apparently that nonrelativistic limit for distribution given by $\tilde{\rho}_e$ also exists. It means that Schrödinger equation doesn't take into account some features of one-electron system that is given by the second distribution. In addition to this it's clear that both distributions enter in formula Eq. (8) on equal basis as if the whole electron is split into two subsystems. Furthermore formulas Eq. (6) and Eq. (7) show that the total energy of the electron is split between these two subsystems in a way as one of the subsystems contain whole mass of the electron. The energy of interaction between these two subsystems is zero (that means they don't have any force interaction) and binding energy distributed uniformly between them. The features listed above can be explained if we admit that an electron obeying Dirac equation represents two particle structure where first particle is a zero electric charge mass of electron and the second one is a zero mass charge of electron. The immediate consequence of this admission is that velocity of the second particle amounts to the $c$-speed of light in vacuum as for the particle with zero rest mass. Indeed it is well known fact that formal velocity of electron is $c$, as eigen numbers for each component of the operator $c\vec{\alpha}$ (see Eq. (1)) are $\pm c$. The other interesting consequence is the model exposes a mechanism for preventing two electrons from occupying the same quantum state, that is the Pauli Exclusion Principle.

Let's assume that two electrons occupy the same quantum state that is they possess the same wave function. Our goal is to come to contradiction with this statement if we assume that each of these two electrons have two particle structure introduced above. It's natural to admit that if wave functions of two electrons coincide, these two systems stay in rest relatively to each other. To stay in rest means to have velocity amounts to zero in the frame where one of the physical bodies is chosen as a reference body. Let's choose the massless charge particle of the first electron as a reference body of our new frame and look from this perspective onto massless charge particle of the second electron. The operator of the velocity for this particle is $c\vec{\alpha}$, in accordance with our interpretation. Each matrix component of $\vec{\alpha}$ has eigen numbers $\pm c$ (no zero eigen number). In addition to this its projections don't commute, that is $[\alpha_i, \alpha_j] \neq 0$ subject to $i \neq j = 1,2,3$. All this means that there is no eigen vector corresponding to zero eigen number as well as zero eigen number, and accordingly no vector to present the state of rest in the chosen frame. We arrive to contradiction.

**Conclusion**

Two forms of density functional were proposed Eq. (4) and (5). Neither of them gives opportunity to get electronic structure analytically, but it is possible to find out the energy of a system if one extracts from experiment electron mass and charge density distributions (see Eq. (8)). It is feasible to make such type of experiment with atomic H-like ions. In order to explain the very fact of existence of different charge and mass density distributions for an electron, model of split electron was introduced. In this model electric charge and mass of electron behave like two point-like particles and as a consequence possess two different wave functions, each of them is appropriate Dirac spinor. The corner stone of the model is a well-known fact that the formal velocity of relativistic electron amounts to $c$-speed of light in vacuum. The properties of the operator $c\vec{\alpha}$ which shows up in Dirac equation, and is interpreted as velocity operator for one of the particle in the model, provide opportunity to demonstrate that two electrons obeying the model cannot occupy the same quantum state. As a whole the model represents an extension of the interpretation for Dirac equation. Critics of the model should keep in mind a few ways of its discrediting. In order to do this they could challenge the following issues.

- Show experimentally that distribution of charge in hydrogen-like ion deviates from the prediction of the model.
- Demonstrate that all density functional forms provided above are not really density functional.
- Find the exact form of nonrelativistic density functional which has minimum.
- Find the exact form of relativistic density functional and prove that it turns into nonrelativistic one in nonrelativistic limit.
- Provide interpretation for operator $c\vec{\alpha}$ and explain why this physical quantity cannot be observed experimentally.
- Prove Pauli theorem (spin and statistics relationship) in nonrelativistic case.
- Explain physical forces behind the mechanism that keeps electric charge and mass of electron together in the same space location.

It is worth to mention here that the model of split electron necessitates (if validated) the reformulation of QED (quantum electrodynamics). Obviously, current of mass should be omitted in a current-potential term explaining the interaction of electromagnetic and electron-positron fields.